\documentclass[aps,nofootinbib]{revtex4}

\usepackage{graphicx}% Include figure files
\usepackage{dcolumn}% Align table columns on decimal point
\usepackage{bm}% bold math
\usepackage{mathbbol}
\usepackage{amsmath}
\usepackage{amssymb}
\usepackage{axodraw}
\usepackage{epsf}
\usepackage{mathrsfs}
\usepackage{slashed}
\usepackage{stmaryrd}
\usepackage{wasysym}
\usepackage{color}

\newcommand{\be}{\begin{equation}}
\newcommand{\ee}{\end{equation}}
\newcommand{\bea}{\begin{eqnarray}}
\newcommand{\eea}{\end{eqnarray}}
\newcommand{\bml}{\begin{mathletters} \baselineskip 10pt}
\newcommand{\eml}{\baselineskip 12pt \end{mathletters}}

\newcommand{\m}{{\scriptscriptstyle -}}
\newcommand{\p}{{\scriptscriptstyle +}}

\newcommand{\simgeq}{\scriptstyle{\stackrel{>}{\sim}}}

\newcommand{\Ep}{E_{\ssvc{p}}}

\def\lambdabar{\protect\@lambdabar}
\def\@lambdabar{%
\relax \bgroup
\def\@tempa{\hbox{\raise.73\ht0
\hbox to0pt{\kern.2\wd0\vrule width.7\wd0
height.1pt depth.1pt\hss}\box0}}%
\mathchoice{\setbox0\hbox{$\displaystyle\lambda$}\@tempa}%
{\setbox0\hbox{$\textstyle\lambda$}\@tempa}%
{\setbox0\hbox{$\scriptstyle\lambda$}\@tempa}%
{\setbox0\hbox{$\scriptscriptstyle\lambda$}\@tempa}%
\egroup }

\newcommand{\vc}[1]{\mbox{\boldmath$#1$}}

\newcommand{\ssvc}[1]{\mbox{\scriptsize\boldmath$#1$}}

\begin{document}

\title{Extreme field physics and QED}

\author{Thomas Heinzl}\email{theinzl@plymouth.ac.uk}
\author{Anton Ilderton}\email{abilderton@plymouth.ac.uk}

\affiliation{School of Mathematics and Statistics, University of
Plymouth\\
Drake Circus, Plymouth PL4 8AA, UK}

\date{\today}

\begin{abstract}
We give a brief overview of the most important QED effects that can be studied in the presence of extreme fields such as those expected at the Vulcan laser upgraded to a power of 10 Petawatts.
\end{abstract}

%\pacs{}
% PACS, the Physics and Astronomy Classification Scheme.


\maketitle

\section{Introduction}

Since the technological breakthrough of chirped pulse amplification \cite{strickland:1985} the power of optical lasers was increased continuously, presently achievable intensities being around $10^{22}$ W/cm$^2$. These are typical for lasers in the 1 Petawatt class such as the current Vulcan laser at the Rutherford-Appleton laboratory and amount to photon numbers of about $10^{18}$ in a cubic laser wave length. By the correspondence principle (large quantum numbers) one expects these laser beams to be very well described by classical electromagnetic fields. The associated (electric) field strength is of the order of $10^{14}$ V/m. In such a strong laser field electrons are shaken so violently that their velocity approaches the speed of light. One therefore concludes that the interactions between lasers and matter, in particular electrons, become relativistic.

One may then go one step further and ask what would happen if, in the presence of an ultra-intense laser, one would probe distances of the order of the electron Compton wave length, $\lambdabar_C = \hbar/mc$. For length scales of this order one expects quantum effects to be all important, and one has to unify relativity and quantum mechanics in the presence of a strong classical field. There is a theory at hand which does exactly this, namely \textit{strong-field} Quantum Electrodynamics (QED). This is a generalisation of the standard quantum field theory of light and matter, QED, which is at present the most accurately tested theory in physics. The tests in question, however, have almost exclusively been done for weak fields, i.e.\ standard QED. Within a particle physics context the strong field sector has only been addressed in a single experiment, namely E-144 at SLAC (see \cite{bamber:1999} for an overview, and below). As this was utilising the 50 GeV SLAC electron beam to produce back-scattered photons of 30 GeV it clearly was a high-energy experiment. Thus, the strong-field, \textit{low-energy} regime of QED remains untested to date. As will become clear below, ultra-high intensity lasers are a unique tool to explore this uncharted region of the standard model. In the following we will show that an upgrade of Vulcan to a power of 10 Petawatt will be a crucial step in this direction.

\section{Strong-Field QED}

The fundamental QED interaction where a photon $\gamma$ couples to an electron-positron pair ($e^\p e^\m$) is depicted in Fig.~1. The coupling strength is determined by the elementary charge, $e$.

\vspace{-1cm}
\begin{figure}[h]
\begin{center}
\includegraphics[scale=0.22]{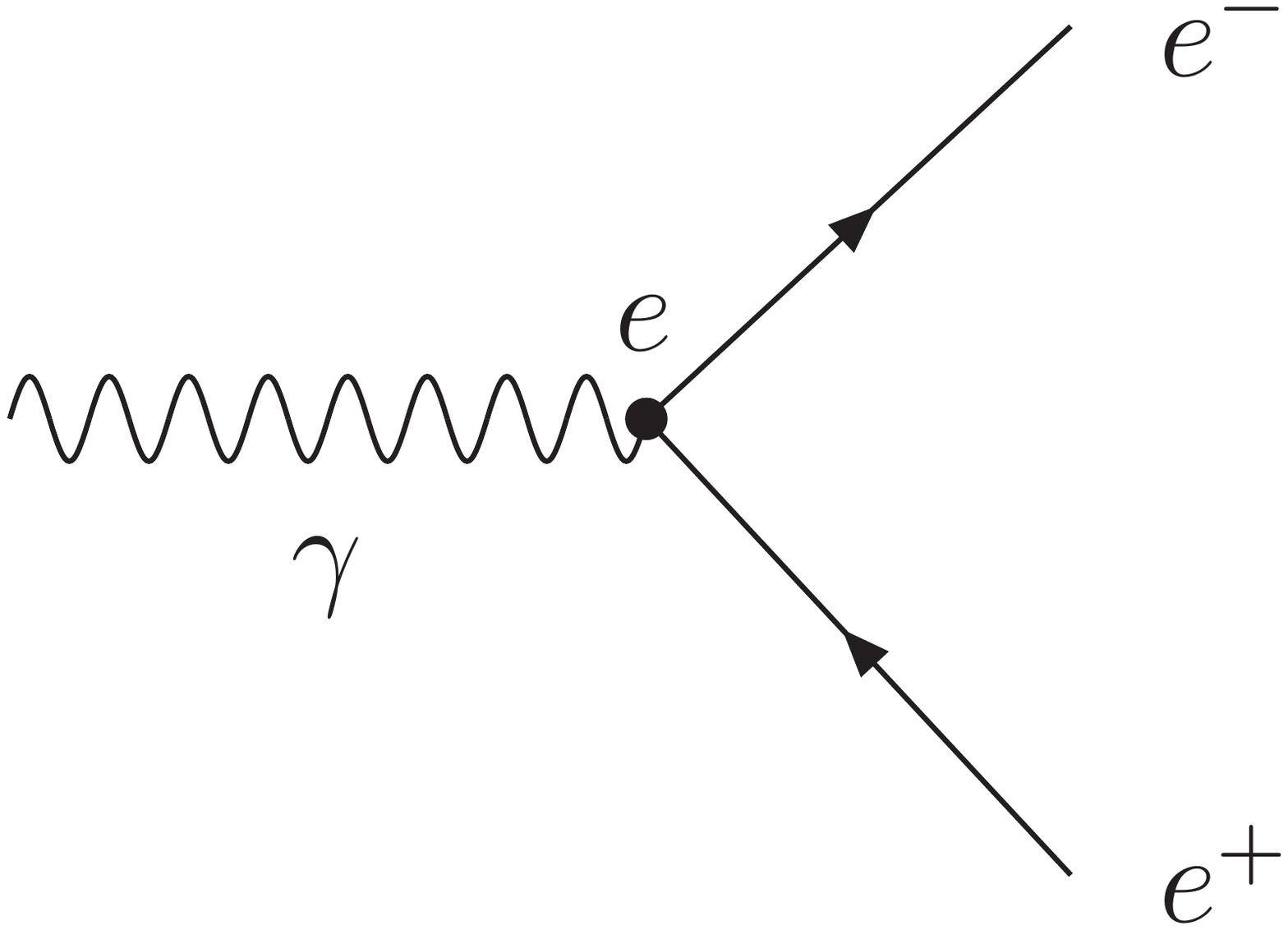}
\end{center}
\vspace{-1cm}
\caption{The elementary interaction in QED.}
\end{figure}

\begin{figure}[h]
\begin{center}
\vspace{-3.8cm}
\includegraphics[scale=0.45,angle=270]{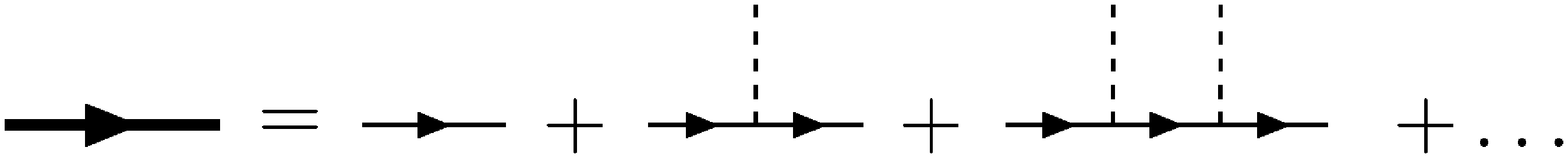}
\end{center}
\vspace{-4.2cm}
\caption{\label{fig:dressedprop}A dressed electron line describing the propagation of a Volkov electron.}
\end{figure}

Viewed as a process involving three real particles on their mass shell the Feynman diagram of Fig.~1 is forbidden by energy-momentum conservation. A massless photon cannot `decay' into an $e^\p e^\m$ pair of total mass $2 m c^2$. Accordingly, if one wants to create matter pairs from light one has to add some additional ingredient, such as extra particles or an external electromagnetic field. For our purposes we imagine that this field is provided by an ultra-high intensity laser. The effect of such a classical background field is to `dress' the fermions which become effective (or quasi) particles with an effective mass $m_* > m$. Theoretically, they may be thought of as being solutions of the Dirac equation in a plane wave (Volkov electrons \cite{volkov:1935}). In Feynman diagrams the electron lines become `fat', as shown in Fig.~\ref{fig:dressedprop}, corresponding to a bare electron absorbing and emitting an arbitrary number of laser photons.

A classical picture of this dressing process is the quiver motion of the electron in the laser field. At high intensities this is so violent that the electron becomes relativistic acquiring a nontrivial $\gamma$-factor that increases its energy and mass. The intensity of the laser beam is measured by a universal `dimensionless laser amplitude',
\be
  a_0 \equiv \frac{e E \lambdabar_L}{m c^2} \; .
\ee
In atomic physics $a_0^{-1}$ is referred to as the Keldysh parameter, as reviewed recently in \cite{dunne:2004}. Obviously, $a_0$ is a purely classical (no $\hbar$ present) ratio of two energies; the energy gain of an electron moving across a laser wave length $\lambdabar_L$ in the field $E$ divided by its rest energy. The numerator does not appear to be Lorentz invariant, but there is indeed an equivalent covariant expression in terms of the energy-momentum tensor of the laser field \cite{Heinzl:2008}. One may say that the quivering electron becomes relativistic when its kinetic energy becomes comparable to its rest energy, i.e.\ when $a_0 \simeq 1$. In Table~\ref{tab:a0} we give an overview of intensities and $a_0$ values achieved or expected at current and future high-power laser facilities. We have used the rule of thumb, $a_0^2 \simeq 5 \times 10^3 P/\mbox{PW}$,  relating $a_0$ to laser power $P$ in Petawatts \cite{mcdonald:1986}.

\begin{table}[h]
\renewcommand{\arraystretch}{1.2}
\caption{\label{tab:a0}Overview of current and future laser facilities: intensities $I$ (in W/cm$^2$) and $a_0$ values (XFEL: X-ray free electron laser at DESY, FZD: Forschungszentrum Dresden-Rossendorf, ELI: Extreme Light Infrastructure project, HiPER: High Power laser Energy Research facility).}
\begin{ruledtabular}
\begin{tabular}{llllll}
& XFEL (`goal') & FZD (150~TW) & Vulcan/POLARIS (1PW)& Vulcan (10PW)& ELI/HiPER
\\
$I$  & $10^{27}$ & $10^{20}$ & $10^{22}$ & $10^{23}$ & $10^{25}$ \\
$a_0$ & 10        & 20   & $70$ & $200$ & $5 \times 10^{3}$ \\
\end{tabular}
\end{ruledtabular}
\end{table}

We mention in passing that for Vulcan (10 PW upgrade) intensities, $a_0 \simeq 200$, muons and pions become relativistic. As a general statement one can say that the region $a_0 \gg 1$ is predominantly accessible by high-power \textit{optical} lasers\footnote{The XFEL is `handicapped' by its small wavelength, ${\lambda_L} \simeq 0.1$ nm.}. In the remainder of this presentation we look at various strong-field QED processes paying particular attention to intensity effects signalled by the appearance of $a_0$.

\section{Nonlinear Compton Scattering}

\begin{figure}[h]
\begin{center}
\vspace{-2cm}
\includegraphics[scale=0.4,angle=0]{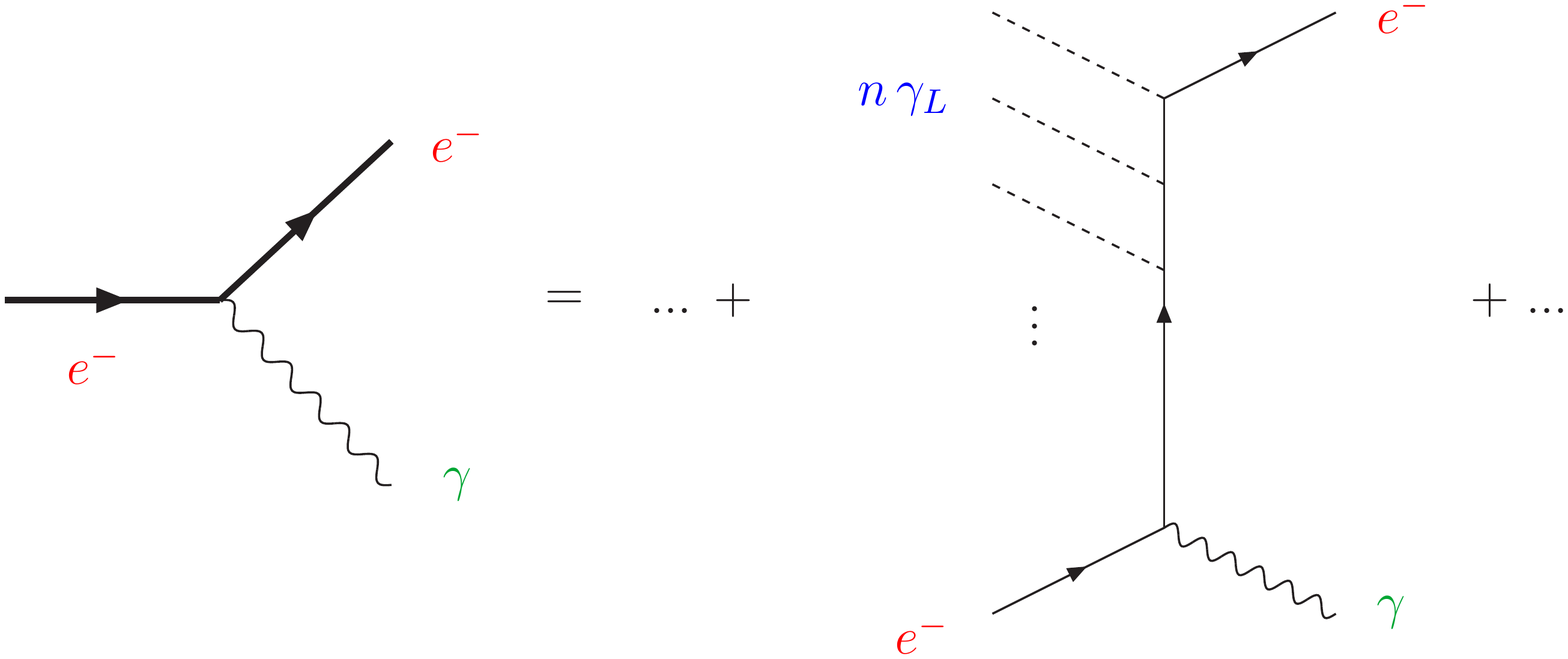}
\vspace{-2cm}
\end{center}
\caption{\label{fig:nlc2L} Feynman diagrams for nonlinear Compton scattering.}
\end{figure}

The process in question is the collision of an electron and a high intensity laser beam such that a photon $\gamma$ is scattered out of the beam. In terms of dressed electrons this is depicted on the left-hand side of Fig.~\ref{fig:nlc2L} which, when expanded in the number of laser photons involved, becomes a sum of diagrams of the type shown on the right-hand side representing the processes
\be \label{NLC}
e + n \gamma_L \to e' + \gamma \; .
\ee
Here, the electron absorbs an arbitrary number $n$ of laser photons (energy $\hbar \omega_L \simeq 1$ eV) before emitting a single photon of energy $\hbar \omega'$.
We may pass from a quantum perspective to a classical electromagnetic wave picture as long as $m c^2$ is the dominant energy scale in the rest frame of the electron. This classical limit is referred to as Thomson scattering. In terms of lab quantities, for the latter to be valid, one requires $\gamma \ll mc^2/\hbar\omega_L \simeq 10^5 \ldots 10^6$ where $\gamma \equiv \Ep /m c^2$ is the $\gamma$-factor of the electrons\footnote{The precise invariant statement is $p \cdot k \ll m^2 c^2$ where $p$ and $k$ are the 4-momenta of incoming electrons and photons, respectively. Note that $k$ is of order $\hbar$, $k = \hbar(\omega_L/c, \vc{k})$.}. It is important to emphasise that the processes (\ref{NLC}) are not suppressed by any threshold effect. Thus, one can study intensity effects at arbitrarily low centre-of-mass energies both for photons and electrons. This is quite a unique feature of nonlinear Thomson or Compton scattering and singles out this process from a particle physics point of view.

\begin{figure}[h]
\begin{center}
%\vspace{-1cm}
\includegraphics[scale=0.5]{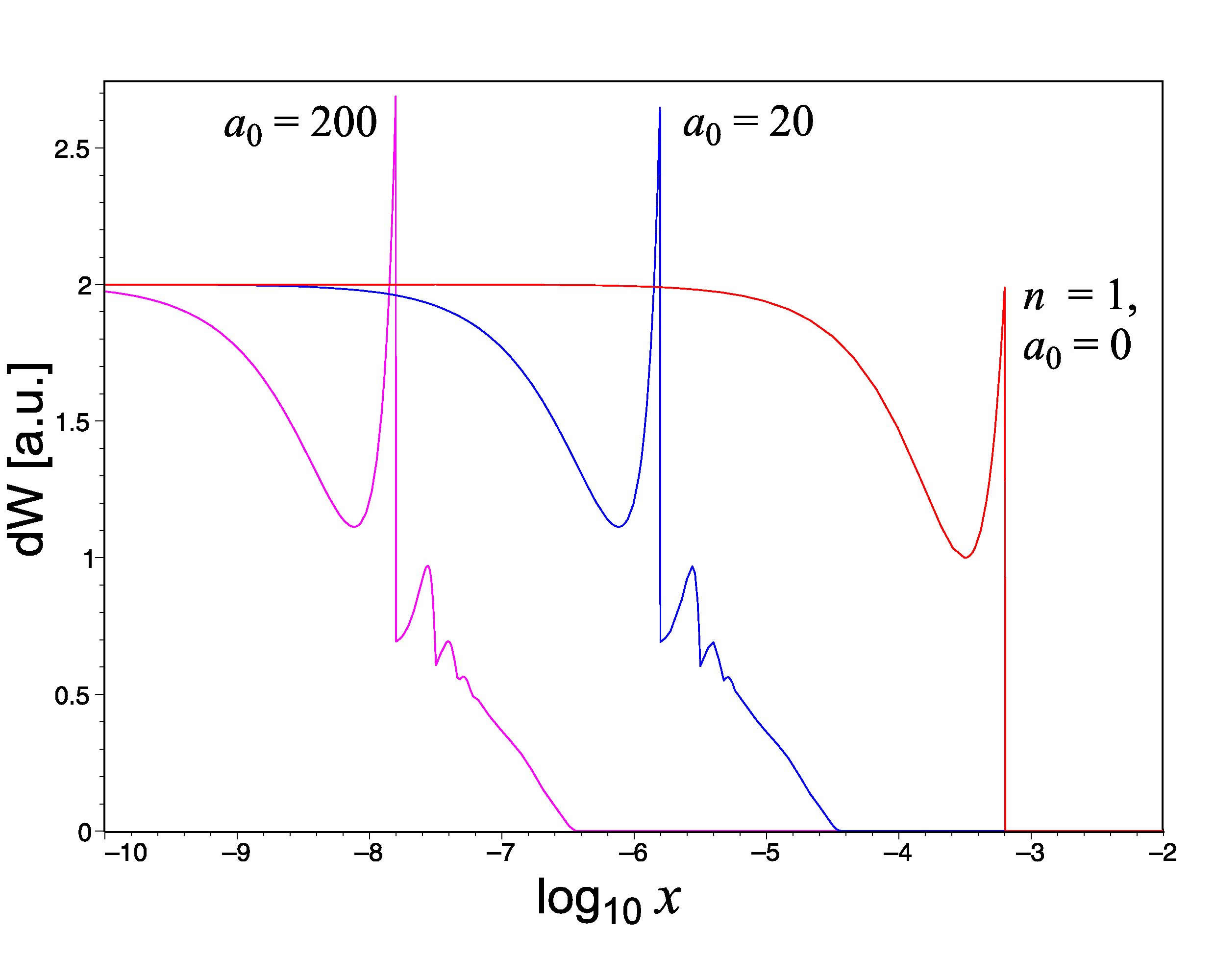}
\end{center}
\vspace{-1cm}
\caption{\label{fig:NLCspectrum} Shift of the (linear) Compton edge as a function of the Lorentz invariant $x$, for electron energy $\Ep = $ 40 MeV and different intensities $a_0$.}
\end{figure}

In Fig.~\ref{fig:NLCspectrum} we show the photon emission rates as a function of a suitably chosen Lorentz invariant, $x$, which basically measures the energy of the scattered photons in any chosen reference frame. The peak at the very right corresponds to standard ($n=1$) low intensity ($a_0 \simeq 0$) Compton back scattering of laser photons colliding with a 40 MeV electron beam. Using a \textit{polarised} laser beam this may be used to produce polarised high energy photons of an energy given by the linear Compton edge value, $\hbar \omega_{\mathrm{max}}^\prime \simeq 4 \gamma^2 \hbar\omega_L$ \cite{Milburn:1962jv,Ballam:1969jr}. Note that, in the lab frame, substantial energy is transferred from electrons to the scattered photons (blue-shift). In an astrophysical context such a process is referred to as `inverse Compton scattering'. This is to be contrasted with `normal' Compton scattering (Compton's original experiment) where the electrons are at rest in the lab and one observes a red-shift of the photon frequency $\omega_L$.

To study intensity effects one uses the (quantum) theory for high-intensity Compton scattering, developed in the 1960's.  This is based on Volkov electrons as asymptotic scattering states \cite{Nikishov:1963,Goldman:1964,Brown:1964} and may be found in the textbook \cite{berestetskii:1982}.
The most striking experimental signal is a red-shift of the linear Compton edge, from $4 \gamma^2 \hbar \omega_L$ to $4 \gamma_*^2 \hbar \omega_L$ with $\gamma_*^2 \equiv \gamma^2/(1+a_0^2)$.  This may be understood in terms of the electron mass shift \cite{Sengupta:1949} mentioned earlier,
\be
  m_* = m \, \sqrt{1 + a_0^2}  \; .
\ee
As the electron `gains weight' ($m \to m_*$) it will recoil  less, reducing the energy transfer to the final state photon, hence the red-shift in the maximum photon energy. This effect is illustrated in the photon spectrum of Fig.~\ref{fig:NLCspectrum} for $a_0 = 20$ and $a_0 = 200$, the latter value expected for the Vulcan 10 PW upgrade. It is interesting to consider what one would observe in the lab frame. We have seen that back-scattering off high-energy electrons ($\gamma \gg 1$) produces a blue-shift (`inverse' Compton). On the other hand, high intensity ($a_0 \gg 1$) produces a red-shift, hence works in the opposite direction. It turns out that there is exact balance in the centre-of-mass frame of the Volkov electrons and the $n$ laser photons, that is when $4 \gamma_*^2 \simeq 1$. This can obviously be achieved by fine-tuning $\gamma$ and $a_0$: for 40 MeV electrons the associated $a_0$ turns out just to be 200. Hence, for $a_0$ of this order or larger one expects an overall red-shift, $\omega' < \omega_L$, as the Volkov electron has become so heavy that it appears almost `static' from the photons' point of view.

The dominant spikes in Fig.~\ref{fig:NLCspectrum} correspond to single-photon absorption, $n=1$ in (\ref{NLC}). However, these spectra also show further peaks corresponding to absorption of $n = 2, 3, \ldots$ laser photons. In the laser community this effect is called higher harmonic generation. Their identification will depend crucially on the size of the background which may wash out the signals of higher harmonics. It will hence be important to simulate the scattering process numerically using realistic beam shapes.

Both the red-shift and the relative width $\delta_w \equiv \Delta \omega'/\omega'$ of the $n=1$ spike depend sensitively on $a_0$, for $a_0 \gg 1$ approximately as $\omega_{\mathrm{max}}^\prime (a_0)/ \omega_{\mathrm{max}}^\prime (0) \simeq 1/a_0^2$ and $\delta_w \sim a_0$. Hence by tuning the initial electron energy $\Ep$ and $a_0$ one can try to design radiation of a particular frequency $\omega'$ and width. For example, using 40 MeV electrons and a laser with $a_0 = 20$ one obtains $\omega' \simeq 0.1$ keV (hard XUV) and a relative width $\delta_w  \simeq 20$ \%. It would be interesting to study how this radiation can be made more mono-energetic.

Nonlinear Compton scattering (\ref{NLC}) has been observed and analysed in the SLAC E-144 experiment \cite{bula:1996,bamber:1999} using 47 GeV electrons from the SLAC beam and a Terawatt laser with $a_0 \simeq 0.4$. This was a high energy ($\gamma \simeq 10^5$) and low intensity ($a_0 < 1$) experiment (hence deep in the `inverse' Compton regime). Photon spectra were not recorded and hence no red-shift was observed \cite{McDonald:1999et}. We reemphasize that this easily accessible process should be studied to a high precision with high-power optical lasers, thus exploring the uncharted region of the standard model of low energies and high intensities -- see also Section~VI.

\section{Pair Production}

\subsection{Stimulated pair production}

\begin{figure}[h]
\begin{center}
\vspace{-2cm}
\includegraphics[scale=0.4, angle=0]{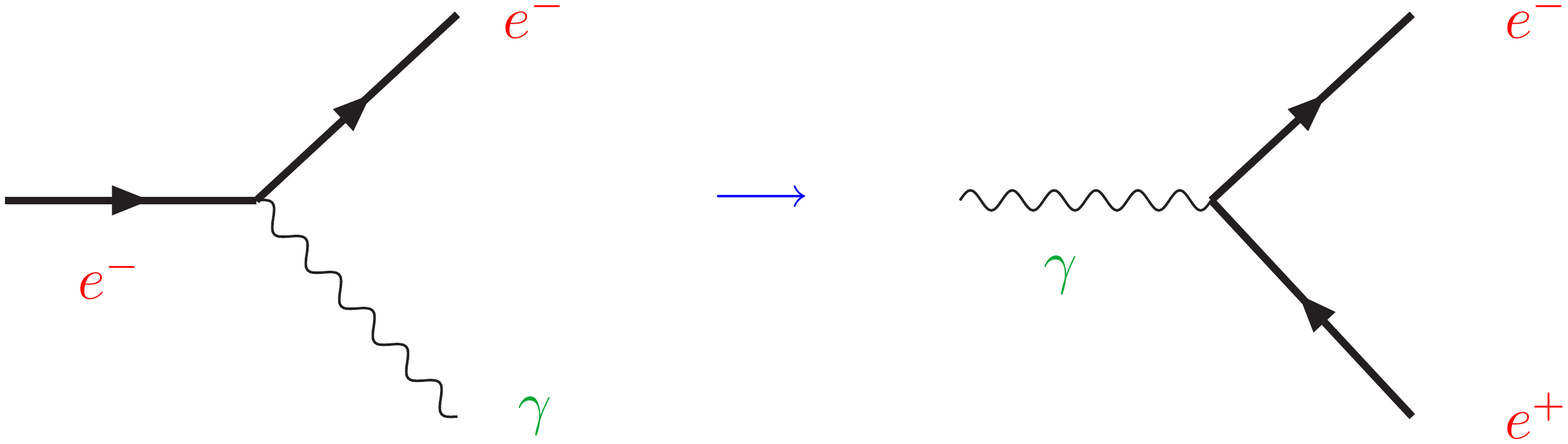}
\end{center}
\vspace{-3cm}
\caption{\label{fig:crossing}Laser assisted pair production, formally obtained from nonlinear Compton scattering via crossing.}
\end{figure}

The next process we will consider is pair production in the presence of an external (laser) field \cite{Reiss:1962}. It is related to nonlinear Compton scattering (\ref{NLC}) via crossing symmetry. In the language of Feynman diagrams this amounts to exchanging, in Fig.~\ref{fig:nlc2L}, the incoming electron with the outgoing photon line as illustrated in Fig.~\ref{fig:crossing}.

\begin{figure}[h]
\begin{center}
\vspace{-1cm}
\includegraphics[scale=0.25,angle=0]{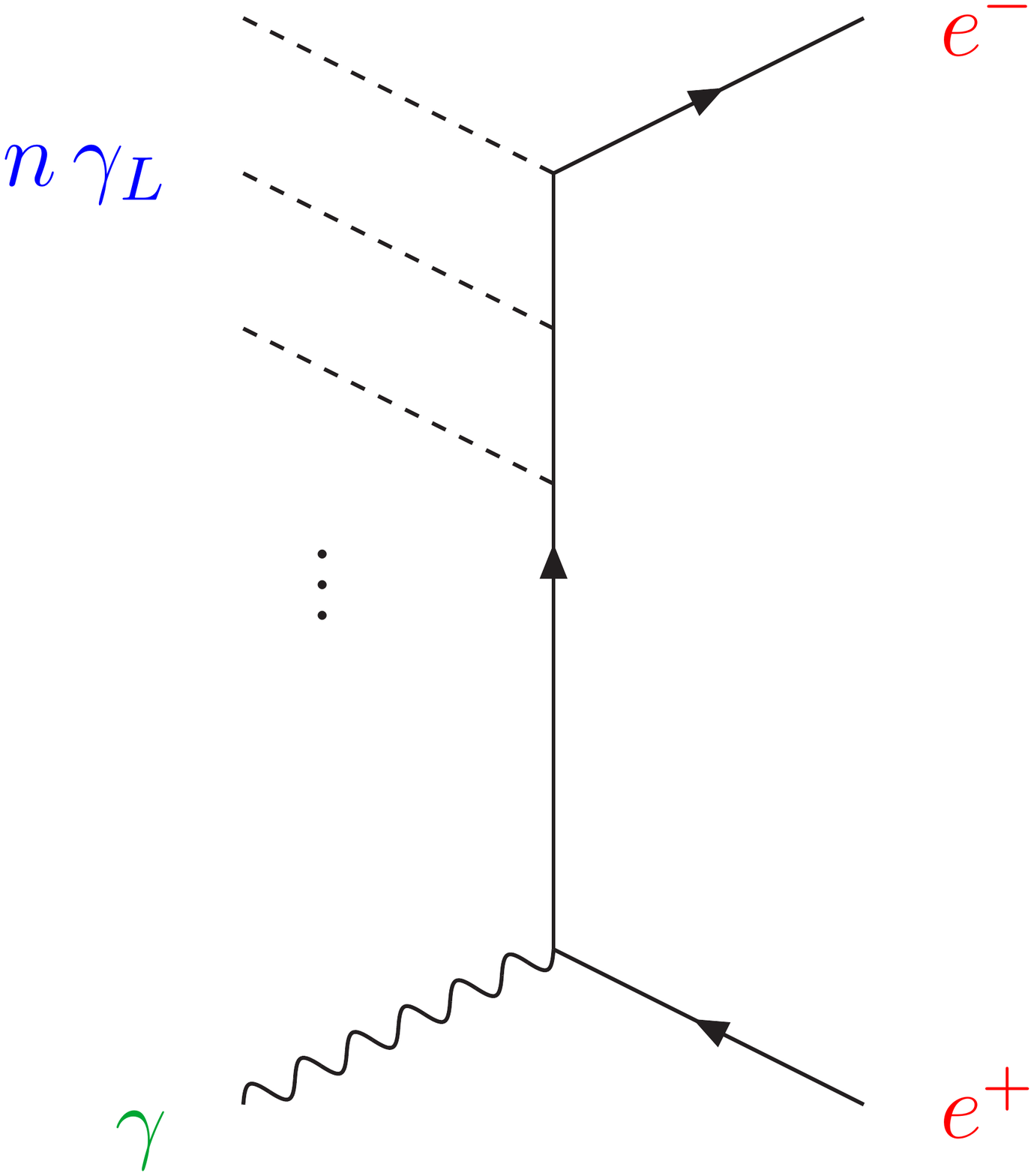}
\vspace{-1.5cm}
\end{center}
\caption{\label{fig:multiBW}Stimulated multi-photon Breit-Wheeler pair production.}
\end{figure}

We are thus summing over all processes $\gamma + n \gamma_L \to e^\p e^\m$ where a photon $\gamma$ interacts with $n$ laser photons and stimulates the production of an electron positron pair (see Fig.~\ref{fig:multiBW}), sometimes called multi-photon Breit-Wheeler pair production \cite{breit:1934}. Unlike (\ref{NLC}) this process has a threshold as the energy $2 m_* c^2 \gtrsim 1$ MeV has to be provided by the initial particles. Note that the effective mass $m_*$ appears here as the pairs are produced within the laser beam. This implies that high intensity actually makes pair creation harder as the threshold is increased by a factor $a_0$ \cite{bamber:1999}. Thus, it is preferable to use high energy photons for pair creation in collisions with lasers of moderate intensity. Again, this was precisely the setup of SLAC E-144 where photons of about 30 GeV were produced via nonlinear Compton (back)scattering (\ref{NLC}) and subsequently used to create pairs according to Fig.~\ref{fig:multiBW} \cite{bamber:1999,burke:1997}.

In Feynman perturbation theory, upon counting the number of photons in Fig.~\ref{fig:multiBW}, one expects the associated production rate (amplitude squared) to go like $W_n \sim \alpha^{n+1} a_0^{2n}$ for $a_0 \ll 1$ \cite{bamber:1999}, $\alpha = 1/137$ being the fine-structure constant. For the $n$th subprocess to happen there is a threshold in photon number given by $n_0 = m_*^2 c^4/\hbar^2\omega\omega_L$ (assuming backscattering). This was indeed confirmed by the SLAC experiment where the kinematics implied $n_0 = 5$. On the other hand, within an all-optical setup one would require an astronomical number of photons of order $n_0 \simeq 10^{12} a_0^2$. Hence, in perturbation theory, the lowest order rate contributing is the corresponding $W_{n_0} \sim \alpha^{n_0+1}$ which is basically zero by power counting in $\alpha$. However, for large $a_0 \gg 1$ perturbation theory is no longer valid and the power counting arguments do not apply. Instead one has to use  nonperturbative expressions like those corresponding to the diagrams in Fig.~\ref{fig:crossing}.  An analogous statement applies to the subject of the next subsection.

\subsection{Spontaneous pair production}

The most spectacular effect in strong-field QED is probably \textit{spontaneous} vacuum pair production as first predicted by Sauter \cite{sauter:1931} and worked out in detail by Schwinger \cite{schwinger:1951}. To see what is involved let us equate the rest energy of an electron with its electromagnetic energy gain upon traversing a Compton wave length $\lambdabar_C$ in an electric field $E$, $m c^2 = e E \lambdabar_C$. This defines the critical field strength
\be \label{EC}
  E_c \equiv \frac{m^2 c^3}{e \hbar} = 1.3 \times 10^{18} \; \mbox{V/m} \; ,
\ee
which translates into a critical intensity of $I_c = 4 \times 10^{29}$ W/cm$^2$ or $a_{0,c} = \lambdabar_L/\lambdabar_C \simeq 10^6$. In a field of this magnitude it becomes energetically favourable for the vacuum to break down by producing electron positron pairs in order to shield and reduce the externally applied field. Note that (\ref{EC}) contains both the velocity of light, $c$, and Planck's constant, $\hbar$, signalling this is both a relativistic and quantum effect necessitating a QED treatment.

\begin{figure}[h]
\begin{center}
\vspace{-1cm}
\includegraphics[scale=0.25,angle=0]{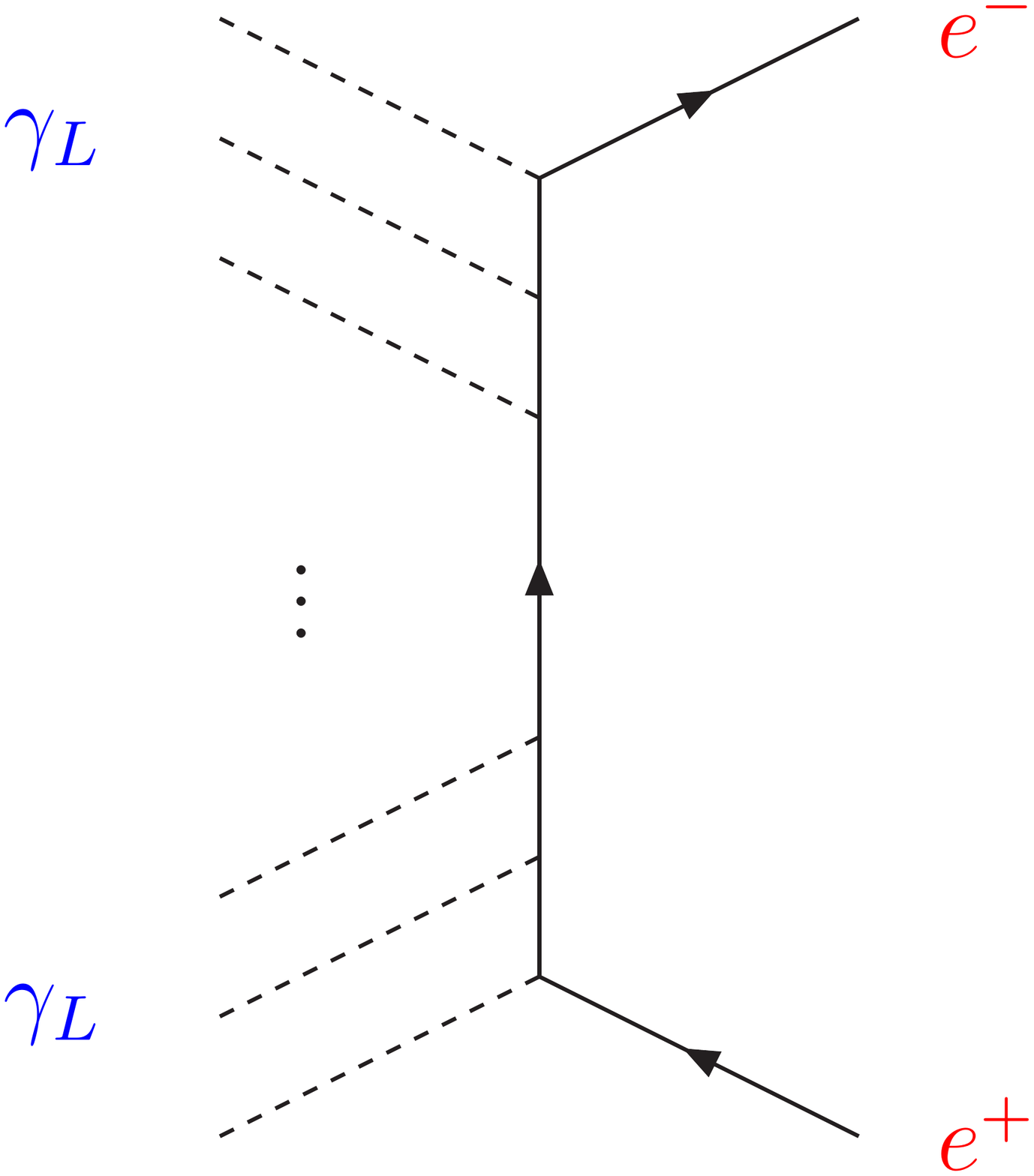}
\vspace{-1.5cm}
\end{center}
\caption{\label{fig:spontPP} Spontaneous vacuum pair production from two counter-propagating laser beams.}
\end{figure}

Energy-momentum conservation rules out spontaneous pair creation by a single laser, hence one needs two (say, counter-propagating) beams to `boil the vacuum' \cite{Ringwald:2003iv}. In Fig.~\ref{fig:spontPP} we have depicted one of infinitely many processes that contribute and (in principle) need to be resummed to obtain the total answer\footnote{Technically, one proceeds in a more elegant fashion by calculating the effective action \protect\cite{schwinger:1951}.}. A crude model for the process is the tunneling of a positron from the Dirac sea through a potential of order $mc^2/e$. The resulting pair production probability is a typical tunneling factor proportional to $\exp (- \pi E_c/E)$ \cite{sauter:1931,schwinger:1951} and implies an enormous suppression for fields below the critical value. On the other hand, there will be a pre-exponential factor which may be large \cite{Bulanov:2004de}. As the calculations can reliably only be done for idealised field configurations there is room for debate and different predictions. Some authors even find no exponential suppression at all \cite{Blaschke:2005hs,Bell:2008hk}.

\section{Vacuum Polarisation}

\begin{figure}[h]
\begin{center}
\vspace{-3cm}
\includegraphics[scale=0.4,angle=0]{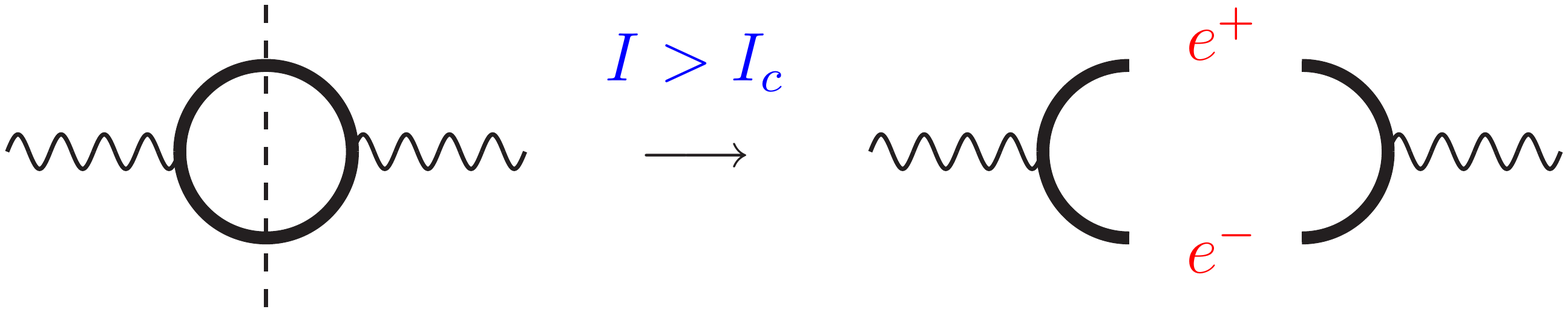}
\end{center}
\vspace{-3.2cm}
\caption{\label{fig:vacpol} Vacuum polarisation with virtual pairs in the loop (left-hand side) breaking up into real pairs above threshold (right-hand side).}
\end{figure}

We have seen above that pairs will be produced once a critical field strength is reached (if not before). Intuitively one may view this process as a break-up of the virtual $e^\p e^\m$ `dipoles' that are omnipresent as fluctuations of the vacuum, see Fig.~\ref{fig:vacpol}. Collectively they produce what is known as vacuum polarisation which is a shielding effect making the physical charge appear reduced as resolution decreases, i.e.\ for large distances. This has observable consequences such as the Lamb shift and the electron and muon anomalous magnetic moments.

In terms of Feynman diagrams vacuum polarisation is depicted as a fermion loop such as on the left-hand side of Fig.~\ref{fig:vacpol}. Both for bare and dressed fermion lines one can associate a mathematical expression with this graph which, above threshold, develops an imaginary part signaling creation of real pairs as discussed in the previous section. As photons `disappear' in this case pair creation is called an \textit{absorptive} process. The real part, on the other hand, describes how virtual pairs polarising the vacuum affect the propagation of probe photons and thus governs all \textit{dispersive} effects. Real and imaginary parts of the vacuum polarisation diagram are related by the optical theorem (or Kramers-Kronig relation) which connects the two sides of Fig.~\ref{fig:vacpol} in a precise quantitative manner.

The most important dispersive effect probably is vacuum birefringence first discussed in Toll's thesis \cite{toll:1952}. The polarised vacuum hence acts as a medium with preferred directions dictated by the external fields. Accordingly, there are two different refractive indices for electromagnetic probe beams of different polarisation states.
These are
\be
  n_\pm = 1 + \frac{\alpha}{45 \pi} (11 \pm 3)  \epsilon^2 + O(\epsilon^4 \nu^2) \; ,
\ee
to lowest order in (dimensionless) intensity $\epsilon^2 \equiv I/I_c \simeq (10^{-6} a_0)^2$ and probe frequency $\nu \equiv \omega/m$. For an X-ray probe of $\omega = 5$ keV and Exawatt class lasers one may achieve values of $\epsilon \simeq \nu \simeq 10^{-2}$.

\begin{figure}[h]
\begin{center}
\includegraphics[scale=0.45,angle=270]{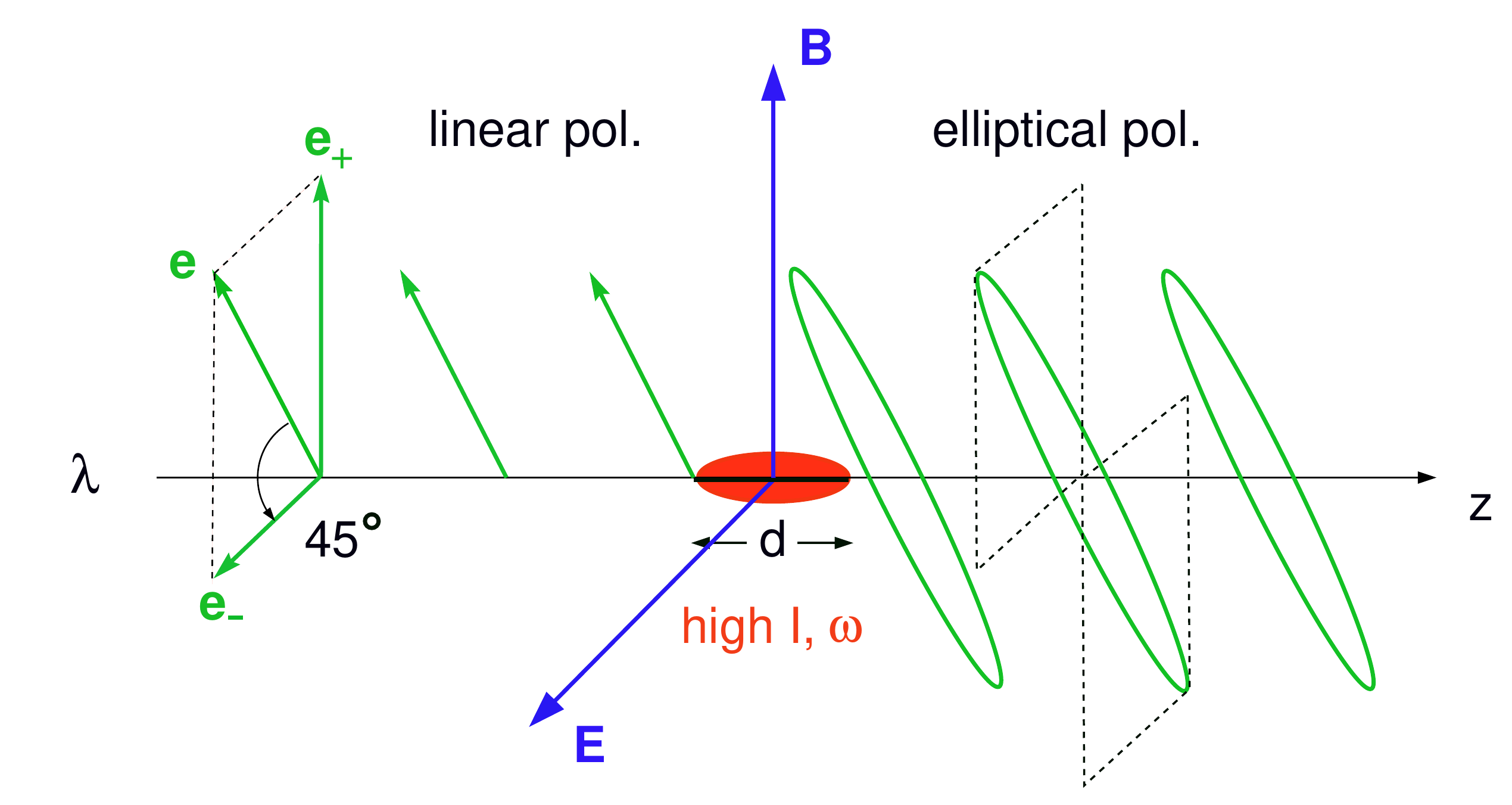}
\end{center}
\caption{\label{fig:polar} Schematic experimental setup to measure vacuum birefringence via an ellipticity signal.}
\end{figure}

The idea \cite{Heinzl:2006xc} is to send a linearly polarised probe beam of sufficiently large frequency $\omega$ into a hot spot of extension $d$ generated by one (or two counter-propagating) laser beams and measure the ellipticity signal, $\delta^2 \sim \omega^2(n_+ - n_-)^2$ caused by a phase retardation of one of the polarisation directions, see Fig.~\ref{fig:polar}.

The leading-order expression for $\delta^2$ is \cite{Heinzl:2006xc}
\be
  \delta^2 = 3.2 \times 10^5 \left( \frac{d}{\mu\mbox{m}} \epsilon^2  \nu \right)^2 \; ,
\ee
and grows quadratically with probe frequency $\nu$, intensity $\epsilon^2$ and spot size $d$ (taken to be the Rayleigh length). Even if all these are  maximised the effect is still extremely small for present and near-future facilities as shown in Table~\ref{tab:ellipticity}.

\begin{table}[h]
\renewcommand{\arraystretch}{1.2}
\caption{\label{tab:ellipticity}Expected ellipticity signals for some high-power laser facilities.}
\begin{ruledtabular}
\begin{tabular}{llll}
facility&  Vulcan/POLARIS (1PW)& Vulcan (10PW)& ELI/HiPER
\\
ellipticity $\delta^2$  & $5 \times 10^{-11}$ & $2 \times 10^{-9}$ & $10^{-7} \ldots 10^{-4}$ \\
\end{tabular}
\end{ruledtabular}
\end{table}

\medskip

X-ray polarimetry is currently sensitive to ellipticities of about $10^{-4}$, the theoretical limit being $10^{-11}$. This requires Exawatt facilities such as ELI or HiPER.
However, if one could produce polarised $\gamma$ beams of MeV energies the signal would go up by several orders of magnitude (with an expansion in $\nu = O(1)$ no longer possible). A possible option is Compton backscattering off the 3 GeV Diamond beam which would yield polarised photons in the 100 MeV range. In this case one could even address questions such as the frequency dependence of the refractive indices where another Kramers-Kronig relation is expected between real and imaginary parts, the presence of the latter being tied to anomalous dispersion, $\partial n /\partial\nu < 0$ \cite{Heinzl:2006pn}.

\begin{figure}[h]
\begin{center}
\includegraphics[scale=0.6,angle=0]{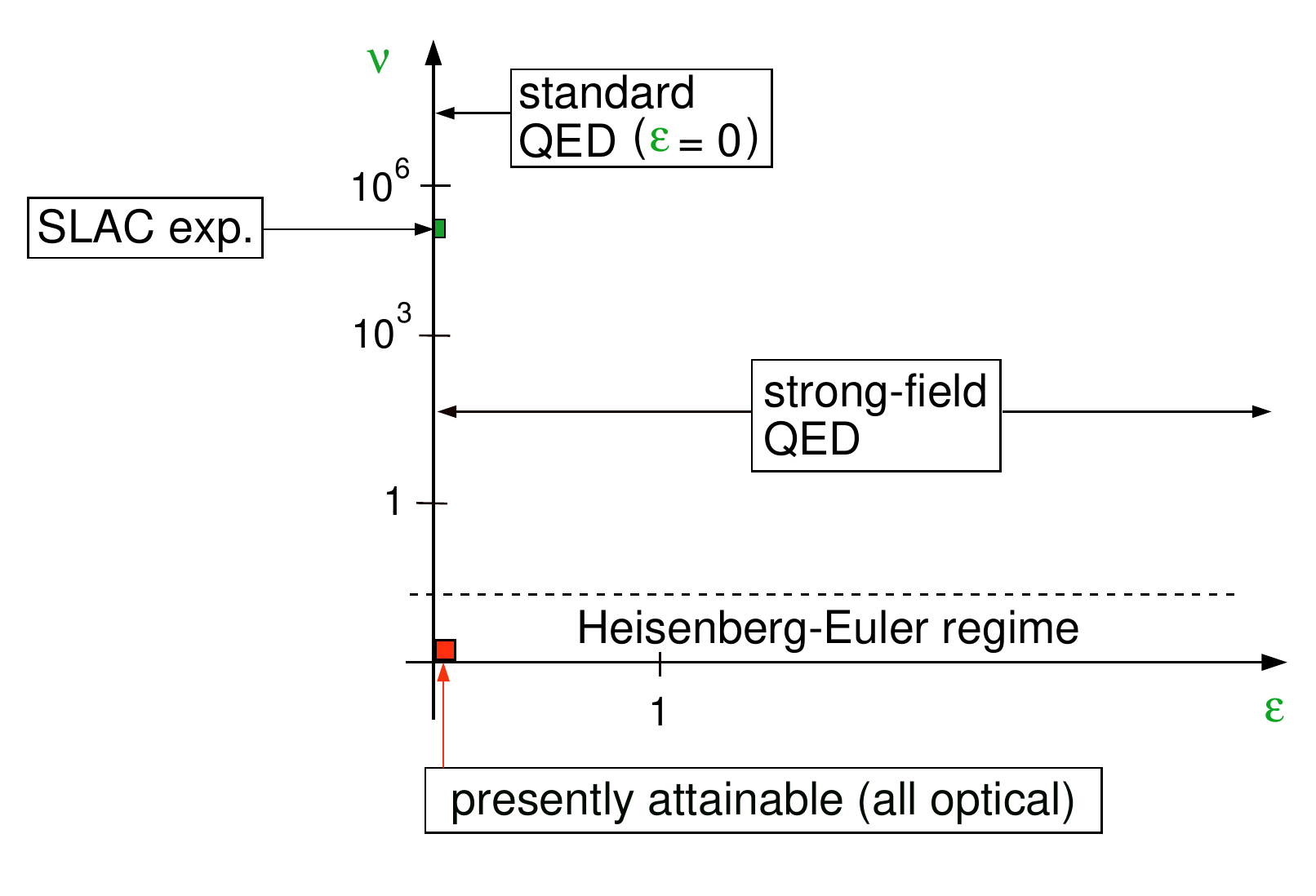}
\vspace{-.5cm}
\end{center}
\caption{\label{fig:parameters} Parameter regime for strong-field QED: energy $\nu$ vs.\ field strength $\epsilon$. High intensities explore the regime to the right of the vertical axis with all-optical experiments staying close to the horizontal one (the Heisenberg-Euler regime).}
\end{figure}

\section{Conclusions}

As illustrated in Fig.~\ref{fig:parameters} high-precision experiments using ultra-intense lasers would be testing an unknown regime of the standard model characterised by extreme field strengths and low energies (at least in an all-optical setup\footnote{Using laser acceleration techniques for charged particles it is possible to reach energies $\Ep \simeq \hbar\omega \, {\simgeq} \, mc^2$, i.e.\ $\nu \, {\simgeq} \, 1$.}). It is not impossible that one could encounter surprises in doing so. One may even discover new weakly interacting sub-eV particles (WISPs) predicted by some theories beyond the standard model \cite{Gies:2007ua}.

In this contribution we have adopted a somewhat more conservative point of view by staying within QED, `only' assuming high intensities. Three areas of immediate interest have been identified, namely (i) nonlinear Compton or Thomson scattering where intensity effects are not suppressed by any threshold, (ii) pair production and (iii) vacuum birefringence. In the latter two cases, there are threshold and/or suppression factors to be overcome. For the Vulcan 10 PW upgrade these are typically of the order of $10^{-6}$ the inverse of which governs the Schwinger exponent. Any ideas on how to reduce the suppression would certainly be highly welcome.

\acknowledgments

We thank Peter Norreys for organising an enjoyable and fruitful meeting.

\bibliographystyle{../../preprints/bibfiles/h_physrev}
\bibliography{../../preprints/bibfiles/laser}

\end{document}